\documentclass[conference]{IEEEtran}
\usepackage[utf8]{inputenc}
\usepackage{graphics}
\usepackage{graphicx}
\usepackage[spaces,hyphens]{url}
\usepackage{algorithm}
\usepackage{algpseudocode}
\usepackage{algorithmicx}
\algrenewcommand\algorithmicrequire{\textbf{Input:}}
\algrenewcommand\algorithmicensure{\textbf{Output:}}
\usepackage{caption}
\usepackage{subcaption}
\usepackage{multirow}
\usepackage{amsthm}
\usepackage{xcolor}
\usepackage{easyReview}
\usepackage{amsmath}
\usepackage{amssymb}

\definecolor{codegreen}{rgb}{0,0.6,0}
\definecolor{codegray}{rgb}{0.5,0.5,0.5}
\definecolor{codepurple}{rgb}{0.58,0,0.82}
\definecolor{backcolour}{rgb}{0.95,0.95,0.92}
\usepackage{listings}
\lstdefinelanguage{tl}{
 morekeywords={ using, for, with, topology, mode, create, of, size, weights, over, uses, limits, use, for,  struct, cpu, component,max,mem, space, net,rate, kbps,ceil, burst, k, nodeLabel, nodeTemplate, artifact, device, located, at, mb, nodes,node, key,components,import,over,actor, colocate,timer,within,msec, pub, MB, rep,app, deploy, separate, on, copies, sec, message'{','}'},
 keywordstyle=\color{magenta},
 backgroundcolor=\color{backcolour},   
 stringstyle=\color{codepurple},
 morecomment=[l]{//}, 
 morecomment=[s]{/*}{*/}, 
 morestring=[b]",
 commentstyle=\color{codegreen},
 numberstyle=\tiny\color{codegray},
 stringstyle=\color{codepurple},
basicstyle=\ttfamily\tiny,
breakatwhitespace=false,         
breaklines=true,                 
captionpos=b,                    
keepspaces=true,                 
numbers=none,                    
numbersep=5pt,                  
showspaces=false,                
showstringspaces=false,
showtabs=false,                  
tabsize=2}

\title{Peer-to-Peer Communication Trade-Offs for Smart Grid Applications}
\author{\IEEEauthorblockN{Purboday Ghosh\IEEEauthorrefmark{1}, Shashank Shekhar\IEEEauthorrefmark{2}, 
Yashen Lin\IEEEauthorrefmark{3},
Ulrich Muenz\IEEEauthorrefmark{2}, Gabor Karsai\IEEEauthorrefmark{1}}
\\ 
\IEEEauthorblockA{\IEEEauthorrefmark{1}Vanderbilt University, Nashville, TN, USA; \IEEEauthorrefmark{2}Siemens Technology, Princeton, NJ, USA 
\\ and 
\IEEEauthorrefmark{3} National Renewable Energy Laboratory (NREL), Golden, CO, USA
}
\\	
\IEEEauthorblockA{Email: \IEEEauthorrefmark{1}\{purboday.ghosh; gabor.karsai\}@vanderbilt.edu, 
\IEEEauthorrefmark{2}\{shashankshekhar; ulrich.muenz\}@siemens.com 
\\ and
\IEEEauthorrefmark{3}yashen.lin@nrel.gov
}
}

\date{June 2021}

\begin{document}

\maketitle
\thispagestyle{plain}
\pagestyle{plain}

\begin{abstract}

Virtual topologies in peer-to-peer networks can reduce the traffic consumed by altering the logical connectivity of peers without altering the underlying network. However, such sparsely connected virtual topologies do not focus on the needs for smart grid applications, which is information dissemination throughout the network, and in turn degrade the performance of distributed control algorithms running on peer-to-peer networks. This paper provides a flexible solution for application developers to prototype and deploy different virtual topologies that balances these trade-offs. First, it introduces a configurable virtual communication topology framework, TopLinkMgr, which enables users to specify any chosen connectivity configuration and deploy peer-to-peer applications using it. Second, it proposes a novel fault-tolerant self-adaptive virtual topology management algorithm, Bounded Path Dissemination, that can ensure the dissemination of information to all peers within a specified threshold. Experiments show that the algorithm improves on convergence speed and accuracy over state-of-the-art methods and is also robust against node failures while consuming significantly less communication bandwidth. 
\end{abstract}

\section{Introduction}


In peer-to-peer architectures deployed for the modern electric grid, a computational \textit{cyber} component with communication capabilities is associated with the various \textit{physical} elements of the grid, such as distributed energy resources (DERs), solar photovoltaic (PV), sensors, and relays. This realizes the \textit{cyber-physical} system, on top of which various energy management schemes are applied. The integration of intelligent algorithms into the electric grid has essentially given rise to the concept of \textit{Smart Grid} \cite{fang2011smart}. 

A peer-to-peer energy management system (EMS) for a smart grid comprises a set of algorithms that achieve a common goal of the system or perform some action. It achieves this through the exchange of information with every other participant without going through a centralized controller. Thus, each component combines its local measurements with the information received from its peers at each iteration to produce an output. The output is usually a type of optimization cost or consensus value (which may be used as a control signal). The output is improved iteratively as more and more information is exchanged. 

The traffic pattern for peer-to-peer EMS agents is always any-to-any. In such cases, a \textit{fully connected} peer-to-peer configuration ensures the highest amount of information dissemination throughout the network, but it also generates the highest network traffic at each round. To be precise, in a network of $n$ nodes, fully connected communication generates $n\times (n-1)$ messages per round. This becomes an issue once the size of the system becomes larger as it might cause messages to be delayed or dropped by the network. One way to overcome this bottleneck is to establish a virtual topology on top of the communication network that creates a sparsely connected communication graph. A virtual topology can be defined as an application-layer overlay that determines how each peer shares information with the rest of the network. Using such virtual topologies, peers can choose a subset of neighbors to share information with instead of flooding the network. However, in such a setup, since fewer information is exchanged at each iteration, the convergence performance of distributed algorithms might be affected both in terms of accuracy and speed. Thus, there exists a trade-off between convergence efficiency and the amount of network traffic generated that must be considered by system designers for peer-to-peer EMS. Application designers need to evaluate their algorithms for different communication connectivity configurations before deciding on the best approach.



To address this need, this paper introduces an integrated, configurable framework, \textit{Topology Manager for Peer-to-Peer Links, or TopLinkMgr}, for implementing and deploying various topology configurations on remote edge devices for evaluation
as well as production deployments. It allows users to specify any custom connectivity graph using a novel text-based domain specific modeling language, \textit{TopLink}.

However, while expert users can use TopLinkMgr to come up with their own topology, manually altering a given topology each time through trial and error is a tedious task for non-expert users. The initial topology that is deployed can often be improved by a few tweaks to the peer-to-peer connectivity without going through a complete redesign. As discussed, the key criteria for a peer-to-peer smart grid application is the effective dissemination of information to all parts of the network. Thus, the final chosen topology should be able to improve on that characteristic.

This brings about a second challenge - how to devise a strategy that can automatically adapt and improve the dissemination of information to all peers in the case of a sparsely connected peer-to-peer communication topology, especially for non-expert users? In terms of graph theory, the problem can be formally defined as, given a set of peers and $V$, determining a set of virtual connections $E: V \times V$, such that there exists a virtual path $P_{ij} \{ e : e\in E \}$ between any two nodes $v_i$ and $v_j$ and the minimum path length satisfies $1 \leq min(n(p_{ij})) \leq th$. Here, $n()$ denotes the cardinal number of a set and $th$ denotes a maximum threshold.

We introduce a novel group-based topology management and peer selection algorithm \textit{Bounded Path Dissemination (BPD)}. It dynamically adapts a given virtual communication topology to limit the minimum path length within an adjustable threshold. For a safety-critical field such as smart grids, fault-tolerance also becomes an important factor. The system must still be able to maintain its operational goal under multiple node failures. Thus, the proposed algorithm has dedicated fault-tolerance protocols that can prevent partitioning of the virtual network if one or more nodes drop out, thus ensuring information dissemination.

There has been considerable work on the topic of peer to peer network overlays for distributed applications such as large-scale file sharing and look up, load balancing and so on \cite{lua2005survey}. The aim of this work is to develop a light-weight virtual topology creation algorithm that ensures reliable delivery of application messages for an any to any traffic pattern.

In summary, the main contributions of this paper are as follows:

\begin{itemize}

    \item \textbf{TopLinkMgr}, a configurable  framework for  implementing  and  deploying  various  topology  configurations over RIAPS platform  that allows the deployment of different peer to peer connectivity configurations.
    
    \item \textbf{Bounded Path Dissemination (BPD)}, a topology management algorithm that improves the speed of information dissemination without flooding the network and is resistant to node failures.

\end{itemize}

The rest of the paper is organized as follows: Section \ref{related_work} describes existing works found in literature. Section \ref{TopLinkMgr} explains the architecture and process flow of the TopLinkMgr framework, section \ref{algorithm} explains the different steps of the BPD algorithm and  section \ref{results} shows the experimental results of applying the algorithm and its effect on peer-to-peer algorithm convergence and fault-tolerance in comparison with some of the alternative methods. Concluding remarks and possible future work are mentioned in Section \ref{conclusion}.

\section{Related Work}
\label{related_work}


There is a substantial amount of literature available in the field of distributed cloud and fog computing on the topic of overlay networks. An overlay is a virtual network built on top of the network layer, supported by its infrastructure that is capable of selectively forwarding packets to peers in the network, in an application-specific way \cite{peterson2007computer}. Various types of overlay topologies exist in the literature, and they can be categorized into two categories, namely \textit{structured} and \textit{unstructured}. Structured overlays are static in nature, where the connections are predefined and remain unchanged throughout the operation lifecycle \cite{castro2002secure}, \cite{dhara2010overview}, \cite{shukla2021towards}. On the other hand, in unstructured peer-to-peer overlays, connections are determined at run-time and can change at each round \cite{ganesh2003peer} , \cite{gencc2005peer}, \cite{jin2010unstructured}. Among them, gossip-based protocols have found widespread adoption in industrial distributed systems use cases such as \cite{lakshman2010cassandra}, \cite{androulaki2018hyperledger}. In a gossip-style communication, at each iteration a peer randomly selects a subset of peers to send its updates to. Although the ideas and concepts used in these approaches can be generalized and have influenced our design, all are focused mainly on the efficient organization of nodes to optimize parameters such as latency \cite{chen2021building}, for file storage and look up \cite{monga2019elfstore}, \cite{shojafar2017flaps}. In \cite{voulgaris2005cyclon}, a peer selection strategy is introduced which is built by improving the randomized shuffling protocol. Using this protocol, an efficient topology management scheme was developed in \cite{jelasity2005t} for a distributed file sharing application. Other algorithms seek to minimize proximity to centralized cloud data centers \cite{costa2020overlay} or dynamic clustering in the case of moving nodes, such as vehicles \cite{rashid2020reliable}. Although the general principles and solution techniques used in these works have inspired this paper, all of them focus on factors important for large-scale dynamic distributed systems where better load balancing and faster response times are desired. For a peer-to-peer smart grid application, the key factor is the distribution of information generated from all peers to all other peers throughout the network efficiently and quickly.

 Overlay solutions that have been implemented recently for smart grid platforms \cite{marzal2017novel}, \cite{tebekaemi2019secure}, \cite{orda2021applying}. These approaches mainly look at using some heuristics to optimize the routing efficiency of messages based on information about the power network. However, the experiments performed always assume a network model that has a few hops to any node, which is the problem that we are trying to solve.
 

Thus, the existing approaches do not solve the performance network usage trade-offs that are important for peer-to-peer smart grid applications, nor do they provide the flexibility for application developers to implement and customize the communication topology for their algorithms.

\section{TopoLogy Manager for Peer-to-Peer Links}
\label{TopLinkMgr}


\begin{figure}[t!]
    \centering
    \includegraphics[scale=0.2]{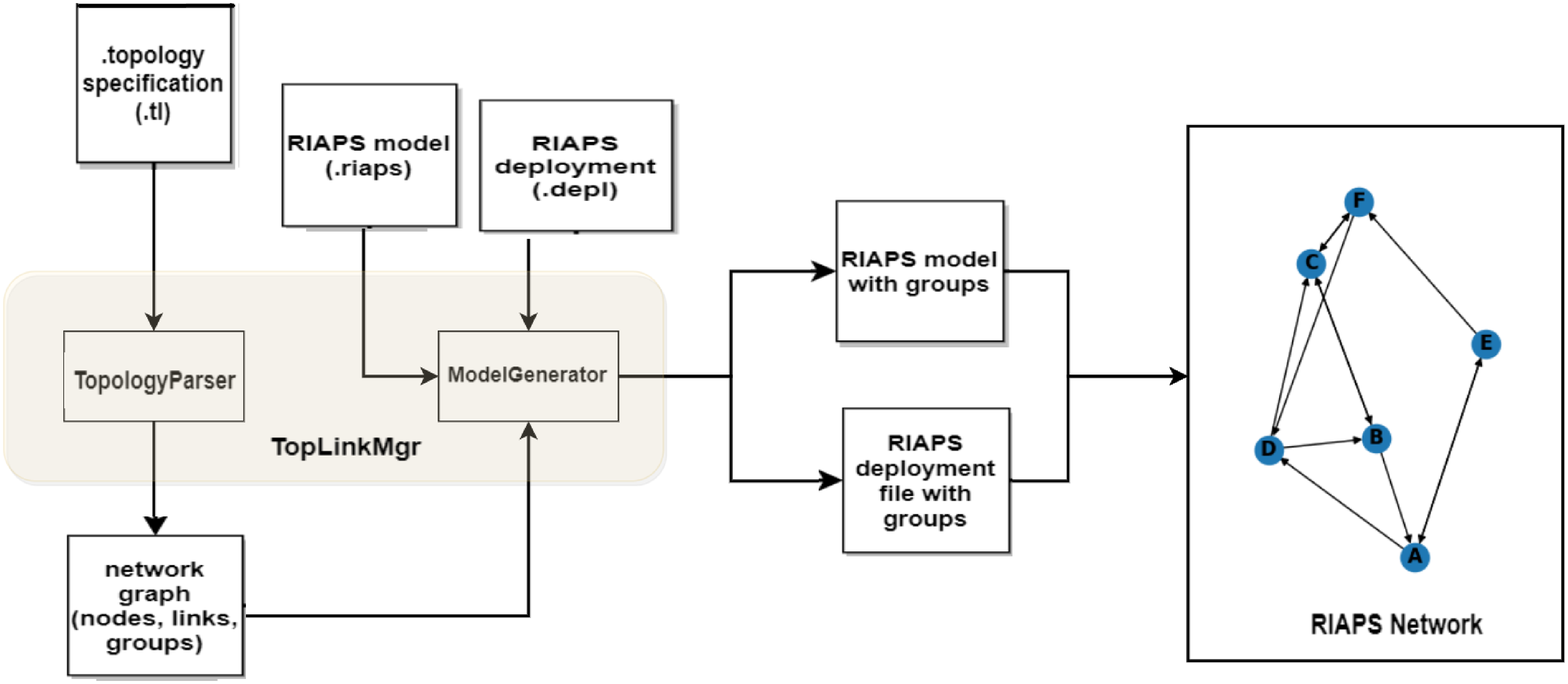}
    \caption{TopLinkMgr Architecture}
    \label{fig:TopLinkMgr}
\end{figure}

This section describes our configurable virtual communication topology framework, \textit{Topology Manager for Peer-to-Peer Links (TopLinkMgr)}. It acts agnostic to the overlying application logic and can be leveraged by application developers readily without worrying about the implementation details. The main goal of TopLinkMgr is to allow application users to deploy distributed peer-to-peer applications using different communication structures to evaluate and determine the best option depending on the application needs. In order to facilitate that, the manager also comes with a library of topologies which are commonly employed, along with providing flexibility to users to design a completely custom topology from scratch. All these options can be provided through the use of a simple text-based user specification language that has been developed for the utility, called \textit{TopLink}. Using \textit{TopLink}, users can specify the names of participating nodes, the type of communication topology, as well as define the individual links for a custom topology in a \texttt{.tl} file, which is then fed to a \textit{TopologyParser} tool that uses the information to translate the user specifications into the appropriate network graph. 

\noindent \textbf{RIAPS Overview}: To develop and deploy the smart grid applications for this article, we use an integrated, decentralized software framework \textit{Resilient Information Architecture Platform for Smart Grids (RIAPS)} \cite{eisele2017riaps}. RIAPS uses a distributed component-based application model with communicating interfaces called ports, using which application developers can realize any peer-to-peer algorithm. It also comes with several platform-level services, such as remote deployment and control, network discovery, time synchronization and fault-tolerance \cite{ghosh2019design}. These platform features
make RIAPS highly suitable as our chosen implementation platform.

The TopLinkMgr framework utilizes one of the features provided by the RIAPS platform called \texttt{groups}, that enables the user-specified virtual topology to be realized into a working RIAPS application. A group defines a dynamic grouping of components at the application level. A group is defined as a group name in the application model which can then have multiple instances defined during deployment. RIAPS groups provide several functions that include:
    
\noindent  \textit{Group communication}: Members of groups can send and receive private messages that are only circulated within that group. Sometimes it also might be required to share information across groups. This can be achieved by forming a separate group comprising all leaders as a cross-group communication channel.

\noindent  \textit{Membership management}: Components can dynamically join or leave a group at any time. RIAPS also provides API-s for detecting and handling membership changes.

\noindent  \textit{Leader election}: RIAPS groups can also perform leader election using a 
RAFT-based algorithm \cite{ongaro2014search}. Members can also communicate directly with the group leader using special methods.

\noindent  \textit{Voting and consensus}: Group members also have the ability to start a voting process on values or actions and gather the final result based on the votes of all other members.

\noindent \textbf{Process Flow}: A virtual topology can be constructed by assigning peers to groups based on how one peer is logically connected to another. This is done by the module \textit{TopologyParser}, which takes as input a RIAPS model file containing the actor, component and message definitions and a topology specification file written using \textit{TopLink}. For each set of communicating links that connect a source peer to any number of destination peers, a group needs to be created. In this context, a group can be defined as a means of establishing the information flow relation between nodes. In TopLinkMgr, there are two categories of groups, a \textit{send group} for outward communication, and a \textit{receive group} for inward communication. The number of groups is determined by the defined virtual topology, and then each node is assigned to be part of these groups based on the virtual connections. Group categories are relative to each node. 

\begin{figure}[t!]
    \centering
    \includegraphics[scale=0.6]{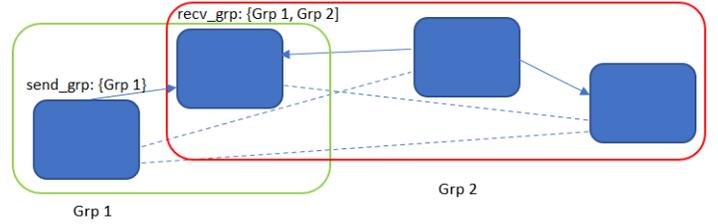}
    \caption{Group formation. The rectangular outlines represent the groups. Then send and receive groups as perceived by the two left-most nodes respectively are illustrated.}
    \label{fig:grps}
\end{figure}

For example, consider the graph of Figure \ref{fig:grps} with four nodes, represented as blue rectangles that are logically connected according to the solid arrows. The direction of the arrows specifies whether the node can send messages to the other node or receive messages from it. The dotted lines represent unused communication links that are not part of the virtual topology. In order to implement this configuration, it would require the formation of two groups, \textit{Grp 1} comprising the left two nodes and \textit{Grp 2} comprising the right three nodes. Since the left most node acts as the sender, TopLinkMgr will assign \textit{Grp 1} as its \textit{send group}. Similarly, the node second from the left receives from the other nodes of both \textit{Grp 1} and \textit{Grp 2}. Thus, ToplInkMgr will assign both groups as its \textit{receive group}.

\begin{figure}[t!]
    \centering
    \begin{subfigure}[t]{0.45\textwidth}
    \centering
\begin{lstlisting}[language=tl]
// RIAPS Application Model
app GroupTestApp {
    // message types
    message SensorData; 
    message NodeData;
   ...
	// Sensor component
    component Sensor(value=1.0) {
      timer clock 10000; 							
      pub sensorReady : SensorData ; // Publish port
    }
    // Averager component
component Averager(Ts=1.0){
      sub sensorReady : SensorData ;	// Subscriber port
      pub thisReady : NodeData; // Publish port
      sub nodeReady : NodeData; // Subscriber port
      ...
    }
...
    // Averager actor
actor Averager(value=0.0,Ts=100.0){
       local SensorData;					// Local message types
       { 
          sensor : Sensor(value=value);		
		  averager : Averager(Ts=Ts);
       }
    }
...
}
}
\end{lstlisting}
   \caption{RIAPS application model}
    \label{fig:modelfile}
    \end{subfigure}
    \begin{subfigure}[t]{0.45\textwidth}
    \centering
\begin{lstlisting}[language=tl]
// topology configuration for RIAPS app
// Actor.Component
for Averager.Averager {
create topology custom; // topology name
over (bbb1, bbb2, bbb3, bbb4, bbb5, bbb6); // nodes
using (bbb1, bbb2), (bbb2, bbb3), (bbb2, bbb5), (bbb2, bbb6), (bbb3, bbb4), (bbb3, bbb6), (bbb4, bbb5), (bbb5, bbb3), (bbb6, bbb1), (bbb6, bbb5) ; // links
inter-group communication on; // turn on communication between group leaders
}
\end{lstlisting}
    \caption{TopLink input file (all features not shown)}
    \label{fig:tl}
    \end{subfigure}
    \caption{TopLinkMgr input files showing an example RIAPS application model and a TopLink specification}
    \label{fig:TopLinkex}
\end{figure}


Finally, once the group assignments are calculated, a RIAPS model deployment file is generated by the \textit{ModelGenerator} utility which also weaves the group configurations and necessary commands into the model file. If a deployment file already exists, the \textit{ModelGenerator} modifies it by adding group arguments, or else it generates a new file. It also automatically generates the required component code to join the appropriate groups. Thus, the output of the entire process flow is a complete working application that implements the specified communication topology among the specified peers. The complete architecture and the process flow are depicted in figure \ref{fig:TopLinkMgr}.

\noindent \textbf{TopLink Specification Language}
\label{TopLink}
TopLink is a text-based domain-specific modeling language that can be used by application developers to describe their chosen topology, either a pre-configured or a completely custom one. It provides special keywords through which users can define different aspects of the topology. Properties that can be defined using the file include:

\begin{itemize}
    \item The RIAPS application actor and component for which the connectivity is being defined.
    \item The type of topology: either preset
    or custom. Currently, the preset topologies that are supported are a ring type and a fixed fan-out randomly generated graph, but in the future we plan to add more.
    \item The node identifiers for the network (IP addresses and host names).
    \item The individual communication links between the nodes (if the topology is custom).
    \item Weights for individual links if the network links are not uniform.
    \item If communication is desired among the leaders of the formed group, it can also be enabled.
\end{itemize}

Figure \ref{fig:TopLinkex} shows a RIAPS application model for a peer-to-peer algorithm that defines the various components, their communication ports and port message types, the actor that encapsulates them, and a corresponding topology specification file that describes a custom topology. Running TopLinkMgr produces a modified model file, a RIAPS deployment file, and the various components' starter code
as its output. The modified model file contains the initialization of the TopLink group and the associated group message type. The deployment file contains the details of how the application actors will be deployed to the various remote nodes (denoted by their IP addresses or their hostnames) and the specific arguments for each actor deployed, if required. The TopLinkMgr generated deployment sets the send and receive group names (which are instances of the TopLink group) as arguments for the corresponding actors based on the topology. Although some features of the language described are based on the RIAPS modeling architecture, the language itself is not tied to RIAPS and can be made to work with other platforms as well using dedicated parsers.


\section{Bounded Path Dissemination Algorithm for Topology Management}
\label{algorithm}

Let us consider a fairly common peer-to-peer use case for smart grids, the distributed optimal power flow (OPF) problem \cite{1033707} for a network of microgrids \cite{bower2014advanced}. OPF is an
optimization problem that minimizes the total generation dispatch cost while satisfying physical and technical constraints on the network \cite{1033707}. Using the primal-dual algorithm to solve the given problem leads to equations \ref{eq:opfprimal} (primal update) and \ref{eq:opfdual} (dual update). In both equations, $k$ stands for the $k$-th node and $t$ for time instant $t$. $x$ and $\hat{\lambda_{t}}$ are the primal and dual variables respectively. $\alpha, \beta, \gamma$ act as weights. In practice, each microgrid control node keeps a local copy of the dual variable. At each iteration, this dual variable is updated both by local measurements as well as communication received from other controllers. This is reflected in the dual update equation \ref{eq:opfdual} where the first term is updated from measurement data and the second term is updated from communication data. Of course, the implementation of such an algorithm requires that the nodes be synchronized (time- or event-based) to drive the iterations. The performance of the algorithm compared to a fully centralized system for sparse communication connectivity becomes worse than all-to-all connectivity.

\begin{equation}
 x_{k,t+1} = Proj_{\chi}[x_{k,t}+\alpha_{k,t}(\nabla f_k(x_{k,t})+A_k\hat{\lambda_{k,t}})]
\label{eq:opfprimal}   
\end{equation} 

\begin{equation}
\hat{\lambda_{t+1}} =  Proj_{\mathbb{R}}[\hat{\lambda_{t}}+\beta_t M(y-b)]-\gamma_t L \hat{\lambda_{t}}
\label{eq:opfdual}
\end{equation}

As discussed previously, the convergence of peer-to-peer smart grid algorithms such as the one discussed above, both in terms of how close the final solution is to an optimal one and how many iterations it takes to reach it, depends on the proper dissemination of information. Thus, any virtual topology deployed must achieve that balance between proper distribution of messages and limiting network traffic. Another important consideration for a safety-critical system such as smart grids is its resilience to unexpected faults that cause crashes of individual nodes and lead to partitioning of the virtual topology.


In order to address these requirements, we introduce the \textit{Bounded Path Dissemination} algorithm. This algorithm improves the dissemination of information to all peers in the network by reducing the hop length of the path from one node to all the other nodes within a threshold, which can be specified as an input parameter. Threshold of 1 implies a fully connected graph. BPD limits the minimum number of iterations required for the message generated by a particular node in the network to reach all its peers. This results in speeding up of the information coverage throughout the network  which effectively improves the performance of a peer-to-peer application algorithm
over the unmodified scenario. Additionally, BPD also has dedicated self-healing capabilities that can ensure that peers remain connected even if one or more drop out due to faults.

The following notations and definitions will be used throughout the rest of this section.

\begin{itemize}
    \item $node_i = (id, send\_grp_i, recv\_grp_i)$ denotes a participating node in the network. It acts as the endpoint for peer-to-peer communication. It consists of an identifier $id$, a sending group object $send\_grp_i$ and a receiving group object $recv\_grp_i$ that realizes the topology.
    
    \item $id$ is a unique identifier for a node. In the case of RIAPS, it is a generated universally unique identifier string (uuid).
    
    \item $grp_m = (id, size, send(), recv())$ denotes a group object. It consists of an identifier for the group, the attribute $size$ that stores the number of members currently present in the group and two methods $send()$ and $recv()$ for sending and receiving messages to and from that group.
    
    \item $send\_grp_i = \{ grp_m.id, weight send() \}$ refers to the sending group object of $node_i$. It contains the id of the group that $node_i$ has joined as a send\_grp. The weight attribute represents the cost incurred for the outgoing links of $node_i$. The default value is $1$. The send and receive operation here refers to invoking them on all groups with ids present in send\_grp.
    
    \item $recv\_grp_i = \{ grp_m.id, recv() \}$ refers to the receiving group object of $node_i$. It contains the ids of the groups that $node_i$ has joined as a recv\_grp. The send and receive operation here refers to invoking them on all groups with ids present in send\_grp.
\end{itemize}

The algorithm comprises two stages, namely \textit{Discover Peers} and \textit{Group Update}. The first stage dynamically gathers information about the various node-to-node paths within the topology and their lengths, while the second stage then uses that information to alter some of the paths to ensure that they all lie within the defined threshold. For our setup, an external agent called TopologyManager was responsible for sending commands to each peer to initiate the algorithm stages. However, the same can be triggered by using an internal clock. For smart grids, the number of nodes is determined by the elements of the grid. The network is closed and private for security. Thus, there is no node churn (unlike cloud-distributed systems \cite{WhatisDi32:online}). Therefore, the algorithm starts with an initial configuration of the nodes generated by the topology framework. However, the discovery phase still ensures generalizability since the path lengths are calculated dynamically without any knowledge of the initial configuration.


\subsection{Stage 1: Discover Peers}
In this stage, all peers share information regarding themselves which is propagated through the topology. This information is used by each node to ascertain and store the minimum paths to all other peers in the network in an internal table. It begins when TopologyManager sends a discoverPeers message to all peers in the network. On receiving the message, each node creates a new message containing its own id, a depth variable initialized to 0, and sends it to each of the groups in its recv\_grp. 
It also puts its recv\_grp id into the message.

On receiving the message, the recipient checks if the group id matches with its send\_grp id. If the depth field is lower than the previously recorded depth of the same node, then it updates the table entry. Next, it increases the depth by the weight associated with that send\_grp and forwards it to its recv\_grp. At the end of this round, each node will have a complete table containing the node ids and the depth corresponding to that node, indicating the path cost. The steps are shown in Algorithm \ref{algo:discopeer}.

\begin{algorithm}
\caption{BPD Stage 1: DiscoverPeers}
\label{algo:discopeer}
\begin{algorithmic}[1]
\Require{Initial $send\_grp_i$ and $recv\_grp_i$ for $node_i$}
\Ensure{$path_i.node_j \leftarrow (\{node_{j} : depth_{ij} \} \forall node_j) $ for $node_i$, $i \neq j$ }
\State \textbf{on event} $DiscoverPeers$ \textbf{do}
\State $msg_i \leftarrow$ \{ id : $node_i.id$, depth: $0$, `grp' : $recv\_grp_i$, `order' : $0$, `weight': $recv\_grp_i.weight$ \}
\State $recv\_grp_i.send(msg_i)$
\State \textbf{on event} $send\_grp_i.recv(msg_j)$ \textbf{do}
\If{$msg_j.grp = send\_grp_i.id$}
\State $msg_j.depth \leftarrow msg_j.depth$ + $msg_j.weight$
\If{$msg_j.depth < path_i.node_j.depth$}
\State $path_i.node_j \leftarrow msg_j$
\State $recv\_grp_i.send(msg_j)$
\EndIf
\EndIf
\end{algorithmic}
\end{algorithm}

\subsection{Stage 2: Group Update} This stage uses the fully fleshed-out table entries from Stage 1 to limit the minimum path cost from all nodes of the network to every other node based on a threshold. This threshold is a parameter that can be decided by the user using the connectivity and the desired information dissemination rate. For the experiments performed in this paper, a threshold was selected using the formula $thresh = (N-1)/2$, where $N$ is the number of nodes in the network, which is half the number of edges required to form a spanning tree, the minimal fully connected graph comprising $N$ nodes. Application designers can select a threshold based on how much they want to compromise convergence performance while trying to reduce bandwidth consumption.
It begins when TopologyManager sends a \textit{groupUpdate} message to all peers in the network. On receiving the message, all nodes look at the paths that they have stored locally. If for any path $depth > threshold$, then that node sends a message containing its id, new depth = 0 and weight to its send\_grp.

On receiving the message, the recipient node increases the depth by the weight of its send\_grp and forwards the message to it. When the message reaches a node for which $depth = thresh - weight$, it appends its send\_grp id to the message and forwards it. When the message reaches the target node, the node reads the group id from the field and joins that group. This effectively means that a new link is added to the network graph linking the target node to an intermediate node between it and the source such that the total path depth from the source does not exceed $thresh$. It terminates when there are no new messages to forward. The steps are shown in Algorithm \ref{algo:grpupdate}.
\begin{algorithm}
\caption{BPD Stage 2: GroupUpdate}
\label{algo:grpupdate}
\begin{algorithmic}[1]
\Require{$path_i$ from Stage 1 for $node_i$, $thresh$}
\Ensure{$path_i.node_j \leftarrow (\{node_{j} : depth_{ij}\ |\ depth_{ij} \leq thresh\} \forall node_j) $ for $node_i$, $i \neq j$ }
\State \textbf{on event} $GroupUpdate$ \textbf{do}
\For{$node_j, depth_j$ in $path_i$}
\If{$depth_j > thresh$}
\State $msg_i \leftarrow$ $\{$`req': $node_i.id$, `id' : $node_j.id$, `length' : $0$, `grp': "", `send\_grp' : $send\_grp_i.id$ $\}$
\State 
\EndIf
\EndFor
\State \textbf{on event} $recv_grp_i.recv(msg_j)$ \textbf{do}
\If{$msg_j.req != node_i.id$}
\If{$msg_j.depth <= thresh - weight_k$ and $msg_j.grp == ``"$}
\State $msg_j.depth \leftarrow msg_j.length + weight_k$
\If{$msg_j.depth == thresh - weight_k$}
\State $msg_j.grp \leftarrow send\_grp_i.id$
\EndIf
\EndIf
\State $send\_grp_i.send(msg_j)$
\Else
\State $joinGroup(msg_j.grp)$
\State $send_grp_i.id \leftarrow msg_j.grp$
\EndIf
\end{algorithmic}
\end{algorithm}

\begin{figure}
    \centering
    \begin{subfigure}[b]{0.3\textwidth}
    \centering
    \includegraphics[width=\textwidth]{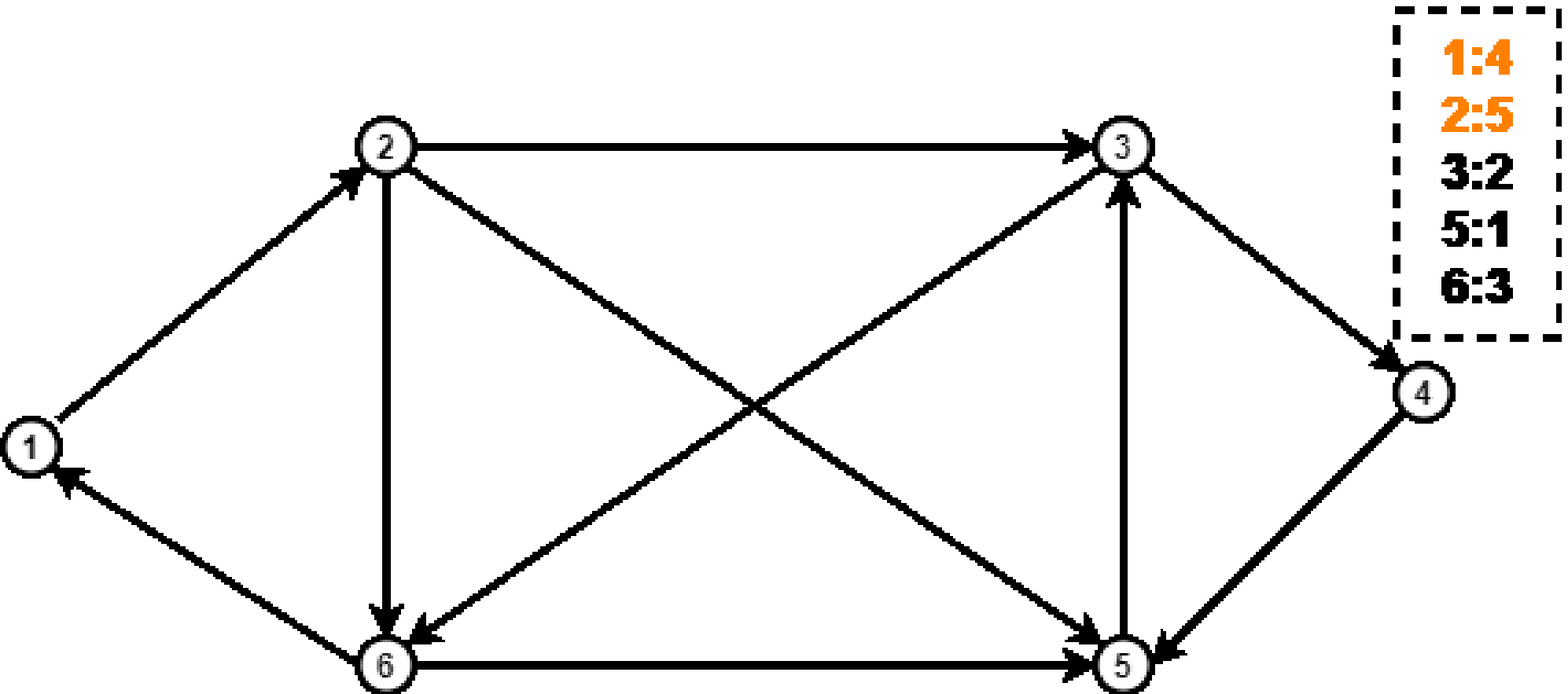}
    \caption{Node 4 after Stage 1}
    \label{fig:stage1}
    \end{subfigure}
    \begin{subfigure}[b]{0.3\textwidth}
    \centering
    \includegraphics[width=\textwidth]{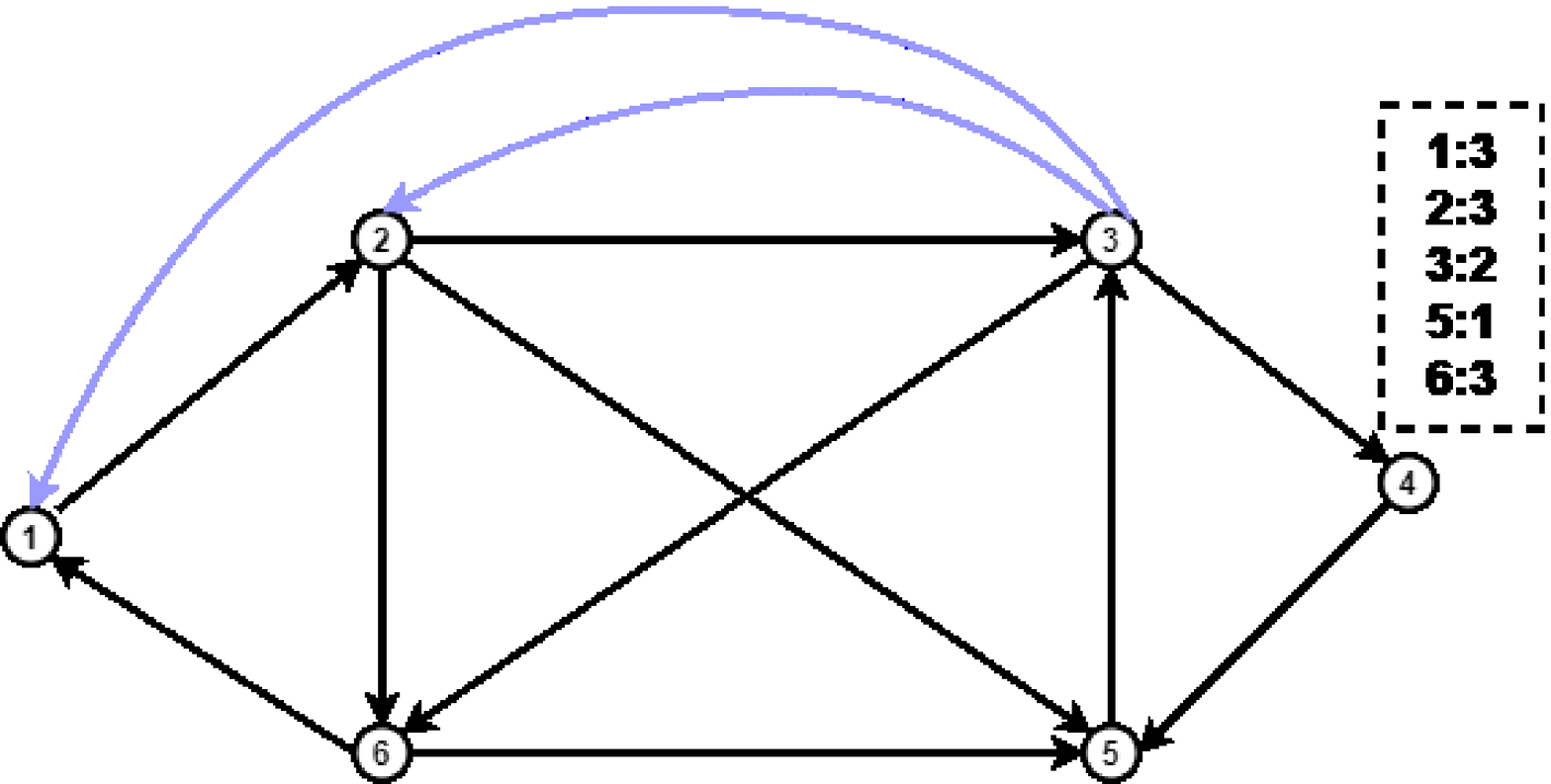}
    \caption{Node 4 after Stage 2}
    \label{fig:stage2}
    \end{subfigure}
    \caption{Bounded Path Dissemination Algorithm on a 6 node topology}
    \label{fig:BPD}
\end{figure}

Figures \ref{fig:stage1} and \ref{fig:stage2} illustrate the working of the two stages of the algorithm for an example 6-node topology. The threshold selected is $3$. In Figure \ref{fig:stage1}, it can be seen that after running the \textit{DiscoverPeers} protocol to determine the relative path lengths for all nodes, for node 4, nodes 1 and 2 are further than the threshold. Thus, when the group update protocol is triggered, nodes 1 and 2 join the send\_grp of node 3, which establishes a path from node 4 to nodes 1 and 2 via node 3. Thus. it can be seen that after completion of the round, the new (shortest) path of nodes 1 and 2 from node 4 becomes 2, which is within the set threshold. Thus, BPD ensures that any message sent from node 4 would reach nodes 1 and 2 within 3 iterations, provided that node 3 is intact. However, that might not always be the case, as unexpected faults might cause nodes to drop out of the network. How BPD can handle such scenarios is discussed in the next subsection. It must be noted that there can be multiple solution topologies that satisfies the requirement. BPD currently does not differentiate between them since in terms of minimum path length they are all equivalent. However, in future works, the algorithm can be optimized, say for e,g., the solution time, or minimum number of edges etc. The algorithm will converge, provided that the starting topology is connected. The speed of convergence depends on both the number of nodes and the threshold selected.


\subsection{Fault Tolerance}
The objective of the fault tolerance protocol is to ensure that the information originating from healthy nodes is distributed to all other peers when one or more nodes drop out due to failure. When such faults happen, it can lead to two things, either some of the path lengths can be altered and exceed the threshold since an intermediate route is removed, or one or more nodes can become isolated or partitioned from the rest of the network. In the first case, the algorithm can self-repair by running the two stages periodically. The period used for the experiments in this paper was $2$ min.
For the second case, there can be two scenarios. The fault-tolerance protocol utilizes the leadership feature of the groups.

\subsubsection{When a member of the send\_grp goes offline} Since for each group, the source node joins it as a send\_grp itself and all other nodes that receive from it join that group as a recv\_grp, if a group has more than one member, it implies that there exists a path from one peer to the other. If a peer is the only member of its send\_grp, it implies that that node is partitioned from the rest of the network. The algorithm then proceeds to connect that node to the rest of the network by joining a new group. If a peer is the lone member in a group, then it sends a \textit{join\_req} message containing its own id and `grp\_type' as `send\_grp' to the group leaders. The leaders respond with a \textit{join\_rep} message by adding to \textit{join\_req} the group id and the group size of the groups in their recv\_grp. On receiving the message, the requesting peer chooses the entry with the minimum size and joins that group as specified in the `grp\_type' field, in this case as a send\_grp. The steps are described in Algorithm \ref{algo:ftsend}

\begin{figure}
    \centering
    \begin{subfigure}[b]{0.3\textwidth}
    \centering
    \includegraphics[width=\textwidth]{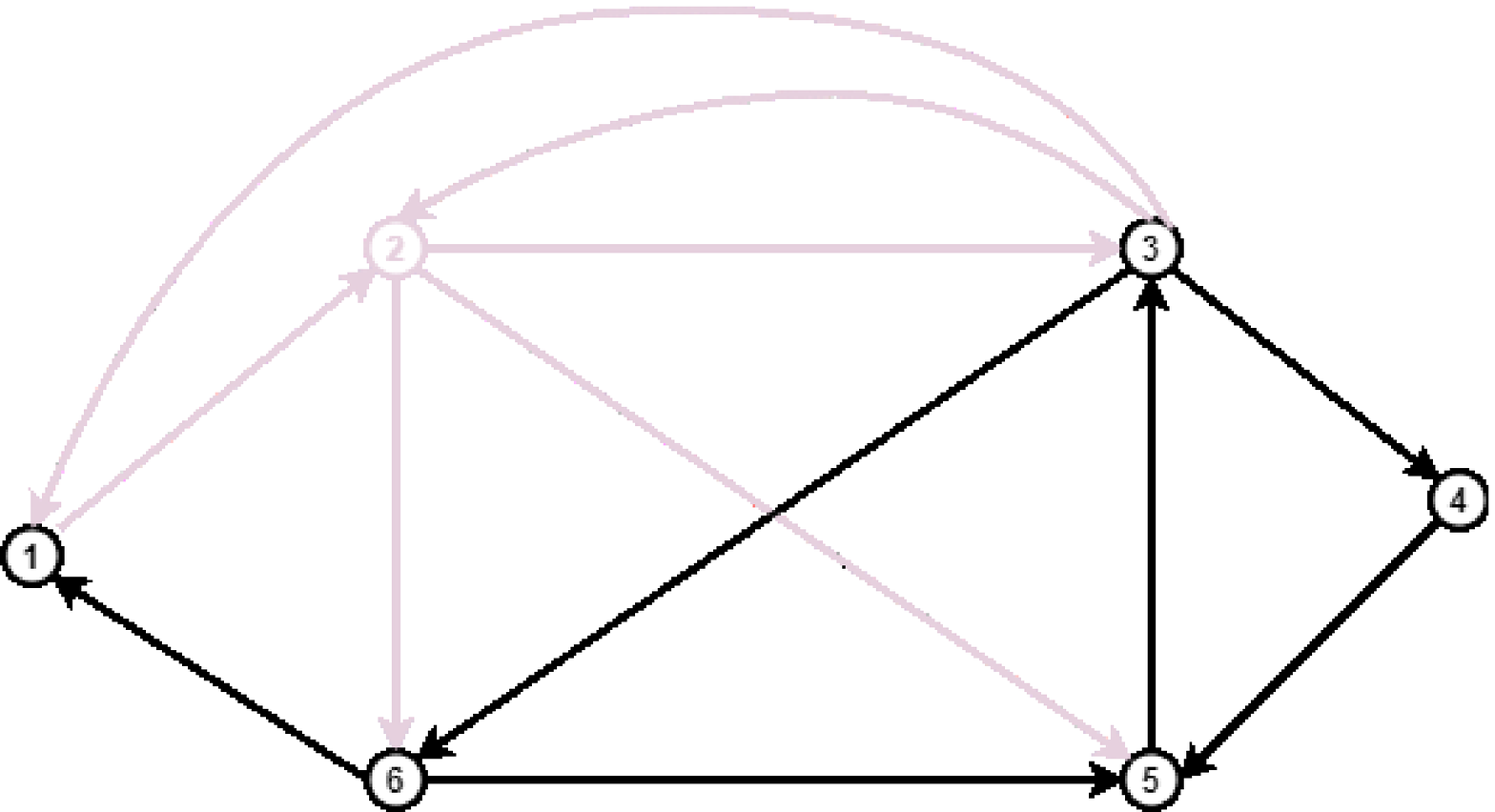}
    \caption{Node 2 goes offline}
    \label{fig:ftstage1}
    \end{subfigure}
    \begin{subfigure}[b]{0.3\textwidth}
    \centering
    \includegraphics[width=\textwidth]{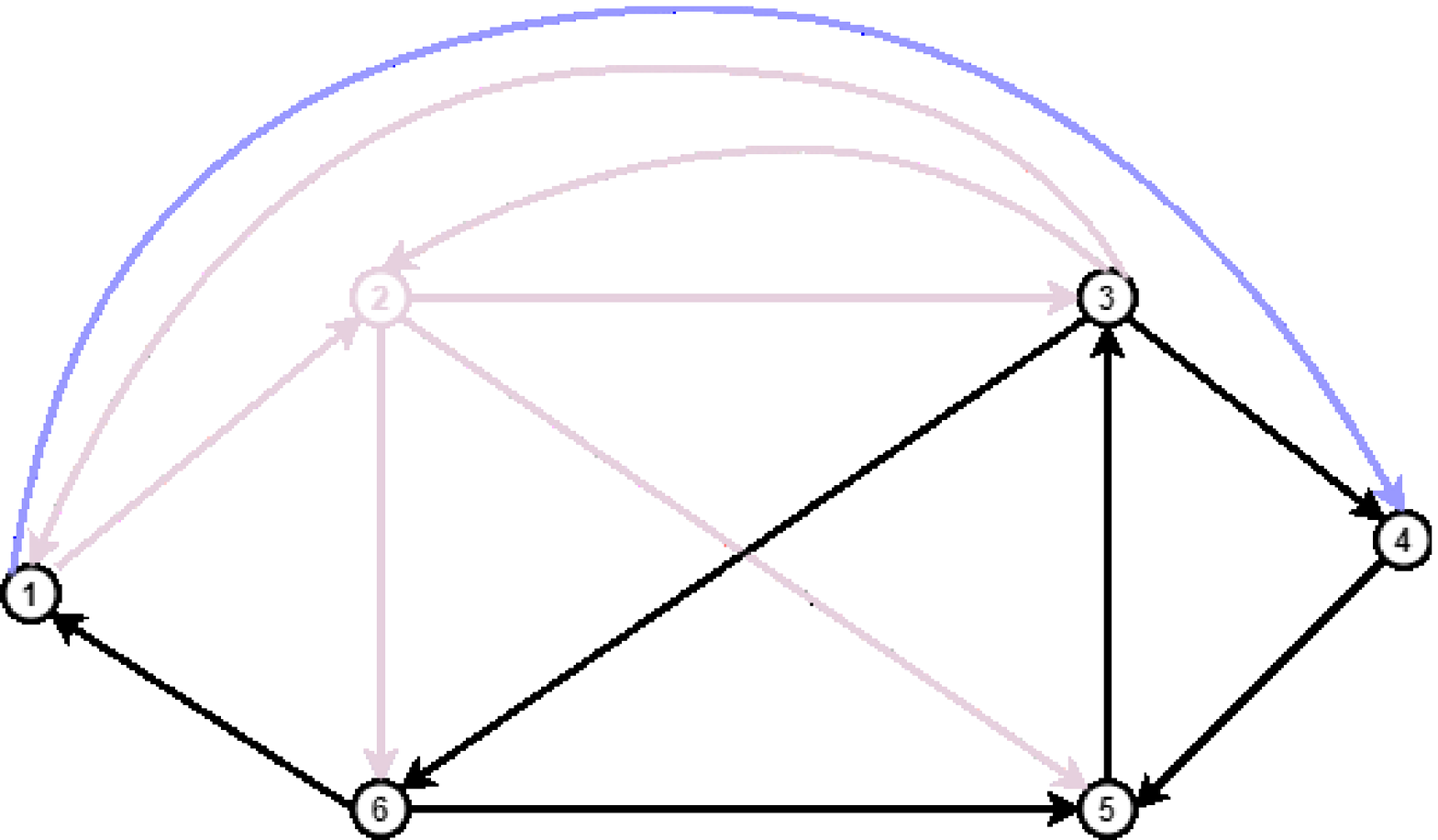}
    \caption{Node 1 is connected to Node 4}
    \label{fig:ftstage2}
    \end{subfigure}
    \caption{Fault-tolerance protocol for the scenario when a member of the send\_grp foes offline}
    \label{fig:BPDFT1}
\end{figure}

In Fig. \ref{fig:BPDFT1}, when node 2 goes offline, node 1 becomes the lone member of its send\_grp, it then initiates the protocol and joins the recv\_grp of node 4, thus reconnecting it.

\begin{algorithm}
\caption{BPD Fault-tolerance: If send\_grp member leaves}
\label{algo:ftsend}
\begin{algorithmic}[1]
\State \textbf{on event} $MemberLeft(group)$ for $node_i$ \textbf{do}
\If{$group.id$ in $send\_grp_i.id$}
\If{$group.size < 2 $}
\State $join\_req_i \leftarrow $ $\{$ `req' : $node_i.id$, `grp\_type' : `send\_grp'$\}$ 
\State $leader\_grp.send(join\_req_i)$
\EndIf
\EndIf
\If{$Memberof(leader\_grp)$}
\State \textbf{on event} $leader\_grp.recv(join\_req_j) \textbf{do}$ 
\State $mingrp \leftarrow minSize(recv\_grp_i)$
\State $join\_rep_j \leftarrow $ $\{$ `req' : $node_i.id$, `grp\_type' : `recv\_grp', `grp': $mingrp.id$, `size' $mingrp.size$ $\}$ 
\EndIf
\State \textbf{on event} $leader\_grp.recv(join\_rep_i)$
\If{$count(join\_rep_i) = leader\_grp.size - 1$}
\State $mingrp \leftarrow minSize(join\_rep_i.grp)$
\State $joinGroup(mingrp)$
\State $send_grp_i.id \leftarrow mingrp.id$
\EndIf
\end{algorithmic}
\end{algorithm}

\subsubsection{When a member of the recv\_grp goes offline} When a node detects a member of the recv\_grp has left, it can imply two possibilities, one, that one of the other receivers in the group has crashed, or two, that the sender node for that group has crashed. For possibility one, the node does not need to act because it still remains connected to its sender, but for possibility two, the node must join a new group. Once a peer detects it, it sends a \textit{grp\_qry} message on that particular group containing the group id. If a recipient has the same group as its send\_grp, it responds with its own id. Once the node receives all the responses, if none of the responses contains the sender's id, it implies that the sender is offline. It then proceeds to send a \textit{join\_req} as described previously but with `grp\_type' as `recv\_grp'. The steps are described in Algorithm \ref{algo:ftrecv}.

\begin{figure}
    \centering
    \begin{subfigure}[b]{0.3\textwidth}
    \centering
    \includegraphics[width=\textwidth]{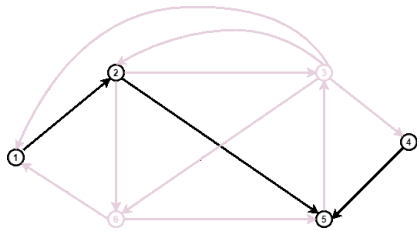}
    \caption{Node 3 and Node 6 go offline}
    \label{fig:ft2stage1}
    \end{subfigure}
    \begin{subfigure}[b]{0.3\textwidth}
    \centering
    \includegraphics[width=\textwidth]{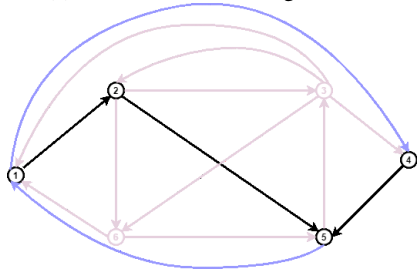}
    \caption{Node 4 and Node 5 are connected to Node 1}
    \label{fig:ft2stage2}
    \end{subfigure}
    \caption{Fault-tolerance protocol for the scenario when a member of the recv\_grp foes offline}
    \label{fig:BPDFT2}
\end{figure}

In Fig. \ref{fig:BPDFT2}, when both node 3 and node 6 go offline, node 4 and node 1 become isolated from the rest of the topology, it then initiates the protocol and node 4 joins the send\_grp of node 1 while node 1 joins the send\_grp of node 5. As a result, all the nonfaulty nodes 1,2,4 and 5 become connected again with both incoming and outgoing routes to each other. In this case, none of the new paths exceeds the threshold distance of 3. However, there might be certain scenarios where that might be the case. In such cases, the next round of the \textit{DiscoverPeers} and the \textit{GroupUpdate} stages can again restructure the connectivity graph to ensure a bounded path.

\begin{algorithm}
\caption{BPD Fault-tolerance: If recv\_grp member leaves}
\label{algo:ftrecv}
\begin{algorithmic}[1]
\State \textbf{on event} $MemberLeft(group)$ for $node_i$ \textbf{do}
\If{$group.id$ in $recv\_grp_i.id$}
\State $grp\_qry_i \leftarrow $ $\{$ `req' : $node_i.id$, `grp\_type' : `recv\_grp', `grp' : $group.id$ $\}$ 
\State $recv\_grp.send(grp\_qry_i)$
\EndIf
\State \textbf{on event} $recv\_grp_i.recv(grp\_qry_j) \textbf{do}$
\If{$grp\_qry_j.grp = send\_grp_i.id$}
\State $grp\_ans_i \leftarrow $ $\{$ `req' : $node_i.id$, `grp\_type' : `recv\_grp', `grp' : $group.id$, `rep' : $node_i.id$ $\}$
\Else
\State  $grp\_ans_i \leftarrow $ $\{$ `req' : $node_i.id$, `grp\_type' : `recv\_grp', `grp' : $group.id$, `rep' : `' $\}$
\EndIf
\State \textbf{on event} $recv\_grp_i.recv(grp\_ans_i)$
\If{$count(grp\_ans_i) = recv\_grp_i.size - 1$}
\If{$grp\_ans_i.rep == `'\ \forall grp\_ans_i$}
\State \textbf{do} Algorithm 3 with grp\_type = `recv\_grp'
\EndIf
\EndIf
\end{algorithmic}
\end{algorithm}

\section{Evaluation and Results}
\label{results}

The experiments performed mainly looked at evaluating two aspects: the performance of a peer-to-peer algorithm in terms of its convergence accuracy and speed as a result of using the algorithm as well as the resilience properties of the algorithm under fault conditions, which were the objectives that the algorithm was designed for. 
\subsection{Experimental setup}
The experiments were carried out using a network of six beaglebones \cite{coley2013beaglebone}  running RIAPS. The communication was carried out using Ethernet. A virtual machine running Linux Ubuntu 18.04 was used as the control node from which the applications were deployed to the target nodes using the RIAPS control graphical user interface.

The algorithm was compared with the base topology (of Fig \ref{algo:discopeer}) without any modification, a fully peer-to-peer approach with all-to-all communication, and a randomized gossip style \textit{pull} pattern that is used in most state-of-the-art schemes. The peer-to-peer algorithm used to perform the experiment was a simple distributed consensus algorithm. For a linear system of $N$ nodes with the local state of the i-th node being denoted as $x_i$, the distributed consensus algorithm updates the state according to the equation \ref{eq:consensus}. For discrete-time, it can be modified to the form of equation \ref{eq: consensus2}, with the derivative being replaced by the difference operator for the discrete case. 

\begin{equation}
    \label{eq:consensus}
    \dot{x_i}(t) = \sum_{j\in N\setminus i}{a_{ij}(x_j(t)-x_i(t))}
\end{equation}

\begin{equation}
\label{eq: consensus2}
    x_i[k+1] = x_i[k] + \sum_{j\in N\setminus i}{a_{ij}(x_j[k]-x_i[k])}
\end{equation}

Here, $a_{ij}$ represents the connectivity between node i and node j. It can be shown that the algorithm theoretically converges to the average of the initial states $x_i[0]$ of the system.

The above algorithm was coded into the Averager RIAPS component logic for the application model described in figure \ref{fig:modelfile} in Python. A logger component was also added to the model to collect data for the experiments. The application was allowed to run for $10$ minutes for each test to ensure sufficient time for the averaging algorithm to converge. The metrics evaluated for algorithm convergence were the percentage deviation from the optimal value (true average) and the number of iterations taken to reach within a $5\%$ tolerance band of that value. The number of messages generated per iteration was also recorded. As expected, all-to-all had the highest with $30$ ($6 \times 5$), followed by Gossip which had $18$ ($3 \times 6$), the graph of Figure \ref{fig:BPD} initially had $10$ links and BPD added $3$ more to make the paths bounded. The initial values for the different for the peer-to-peer averaging algorithm were changed such that it led to a different optimal value for each run of the experiment to eliminate any bias in the results.

\subsection{Algorithm Convergence Results}

\begin{table}
\centering
\begin{tabular}{|p{0.15\columnwidth}|p{0.15\columnwidth}|p{0.15\columnwidth}|p{0.15\columnwidth}|p{0.15\columnwidth}|}
\hline
\textbf{Method}     & \textbf{Deviation from optimal value} & \textbf{Min. number of iterations to reach $\pm 5\%$} & \textbf{Messages per iteration} & \textbf{Time taken (s)} \\ \hline
All-to-All & 4.55\%                       & 445                                    & 30  &  3                  \\ \hline
Gossip     & 10.76\%                      & 172                                    & 18    &    1.8             \\ \hline
Unmodified & 11.32\%                      & 320                                    & 10  & 1                   \\ \hline
BPD        & 6.28\%                       & 60                                     & 13  & 1.3                   \\ \hline
\end{tabular}
\caption{Convergence Performance Comparison}
\label{tab:results1}
\end{table}

Table \ref{tab:results1} lists the convergence performance of the different topology algorithms. The time taken to execute one pass of the control function was $0.01$ ms, averaged across all nodes using the above described setup. In terms of peer selection strategy, as expected, a fully connected network produces the most accurate results with a deviation within $5\%$ of the theoretical optimal value. However, in doing so, it also consumes the most messages per round (iteration), further emphasizing the trade-off that was discussed previously. Bounded Path Dissemination (BPD) improves upon the other approaches on both convergence accuracy and speed. It reaches the steady state the quickest, but the value is not optimal. In our experiments, we also observed sharper fluctuations
in the values as the algorithm progressed through successive rounds compared to the all-to-all configuration. This is because the new information received caused a larger correction in some rounds than in others. Gossip performance lies somewhere in between. Due to its inherently random nature, some nodes perform well, since they can receive all the information, but other nodes might take longer to receive the same. However, Gossip will perform better when a large number of nodes are incoming and outgoing, since the algorithm is automatically scalable.

\subsection{Fault-tolerance Results}

Since the operational efficiency of a peer-to-peer smart grid application depends on the effective dissemination of information to all participating nodes in the network, we measure the success of our fault-tolerance logic on the basis of its ability to maintain network coverage in the presence of faults. Gossip reliability is a measure that has been traditionally used to evaluate broadcast algorithms. It is defined as the percentage of active nodes that can transmit a gossip broadcast, with $100\%$ denoting a perfectly reliable broadcast \cite{leitao2007epidemic}. However, reliability only considers messages originating from one source and forwarded by others until they reach all other nodes. It does not capture the information that shows whether messages originating from all nodes reach all other nodes, which is the case for a peer-to-peer application. Thus, we slightly modify the definition of reliability to be the fraction of messages originating from distinct source nodes that were received by other peers in a peer-to-peer network. We can call this altered reliability the \textit{dissemination efficiency (DE)}. Its value ranges from $0-1$, where $1$ implies that all peers received information generated by every other active member, while $0$ implies that no messages were received. 


Table \ref{tab:ft} shows the average DE for the various techniques under 17\% and 33\% node failure scenarios. It can be seen that the unmodified topology suffers heavily due to the network being partitioned. Gossip is inherently robust since it randomizes the recipients and ensures that faulty nodes are eventually replaced in the sending list. BPD performs similarly to Gossip since it is able to reconnect the graph, thus ensuring that the remaining nodes are able to effectively receive information from each other.

\begin{table}
\centering
\begin{tabular}{|p{0.2\columnwidth}|p{0.1\columnwidth}|p{0.2\columnwidth}|p{0.1\columnwidth}|}
\hline
\textbf{Method}     & \multirow{2}{*}{\textbf{Gossip}} & \multirow{2}{*}{\textbf{Unmodified}} & \multirow{2}{*}{\textbf{BPD}} \\ \cline{1-1}
\textbf{Failure}  &                         &                             &                      \\ \hline
17 $\%$ (1/6)       & 0.82                    & 0.54                        & 0.83                 \\ \hline
33 $\%$ (2/6)       & 0.66                    & 0.17                        & 0.66                 \\ \hline
\end{tabular}
\caption{Dissemination Performance Comparison under Faults}
\label{tab:ft}
\end{table}


\begin{figure}
    \centering
    \includegraphics[width= \columnwidth]{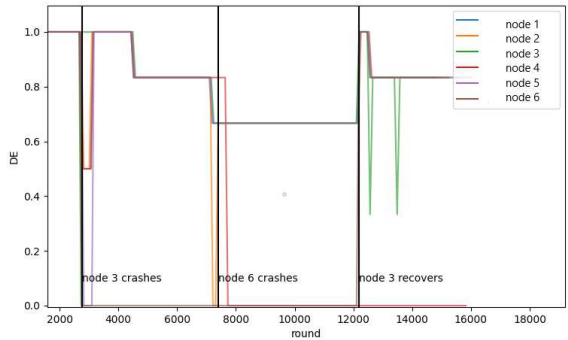}
    \caption{Node-wise Dissemination Efficiency for BPD showing faults and recovery}
    \label{fig:rtft}
\end{figure}

Figure \ref{fig:rtft} shows how the algorithm affects individual nodes in real-time under fault conditions for the starting topology of Fig \ref{fig:stage1}. The plot shows the change in the DE for each node in response to node faults as the number of rounds of message exchange progresses. For the experiments conducted, the period chosen for each round was $10$ ms.
It shows that the DE stabilizes to the expected values of around $0.83 (5/6)$ and $0.667 (4/6)$ after node 3 and node 6 crash respectively 
It also shows that the algorithm is able to restore the connectivity once a faulty node is repaired and it rejoins the network. The average time taken by BPD to restore the topology connectivity following a fault was $16$ ms. There are some cases where it shows the DE to jump up to $1$ right after a fault. This is because to calculate the metric, each node stores a history of the values that it receives and its source, which is refreshed periodically. Thus, it takes some time for the residual entries to be removed, and the true value is then reflected. We will investigate more efficient methods for recording metrics in the future, possibly by adding tags to all outgoing messages or using sequence numbers to differentiate between current and older entries. However, this is only related to the way the data shows up and has no effect on the algorithm performance. 
The sharp downward spikes are due to the time it took for the BPD protocol to be completed after detecting that a group member left. In all our experiments this delay was within the range of $10$ ms, which is quite low compared to the usual sampling time period of peer-to-peer EMS algorithms of $>100$ ms.
 
 \subsection{Network Measurements}
Table \ref{tab:trafficdata} shows the average bandwidth consumed per node and latency data captured for the different methods. As seen from the data, BPD uses about $80\%$ less network bandwidth compared to the fully connected configuration, with the performance of the other two lying in between. This is expected since those two do not take any steps to optimize the virtual topology. However, the methods that employ virtual topologies (BPD modifed and the unmodified) produced a higher latency than the other two. This is due to the fact that using a virtual topology implies that each message needs to go through an additional layer of routing on top of the physical network routing, while in the other two cases it only needs to go through the physical network routing. Although the extra delay is still negligible for it to affect the experiments in this paper, we plan to study it in more detail for larger networks in the future.
 \begin{table}[!h]
\centering
\begin{tabular}{|p{0.2\columnwidth}|p{0.3\columnwidth}|p{0.2\columnwidth}|}
\hline
\textbf{Method} & \textbf{Bandwidth consumed (kB/s)} & \textbf{Latency (ms)} \\ \hline
All-to-All      & 33                            & 1.2                  \\ \hline
Gossip          & 21.5                            & 0.88                  \\ \hline
Unmodified      & 12                            & 3.6                   \\ \hline
BPD             & 6.8                            & 2.4                  \\ \hline
\end{tabular}
\caption{Bandwidth consumed and latency data for the different methods}
\label{tab:trafficdata}
\end{table}
\section{Conclusion}
\label{conclusion}

Peer-to-peer communication plays an important role in a smart grid context in implementing several of the core functionalities. An important question that designers must address is how to achieve a balance between the amount of network traffic generated as the network scales and the preservation of performance goals with respect to the various algorithms that are deployed at each functional layer of the grid. The choice of communication topology is an important design decision that power system engineers must take into account with regard to that trade-off. In this paper, we introduce an integrated framework for smart grid applications that allows users to deploy and prototype different virtual communication topologies for peer-to-peer smart grid applications and empirically evaluate their performance.

We also propose a new algorithm, Bounded Path Dissemination, which ensures the dissemination of information throughout all participating peers in the network within a specified threshold. It also has dedicated fault-tolerance features to prevent partitioning of the virtual topology
if one or more nodes drop out. Experimental evaluations that compare the performance of BPD with other state-of-the-art approaches show improved convergence accuracy and speed for a peer-to-peer application. Similar studies under faulty conditions also show that the algorithm is able to maintain network communication connectivity in the presence of such faults and the Dissemination Efficiency factor is on par with a robust randomized gossip technology. With the integration of more ad hoc participants in a grid such as hybrid electric vehicles, a self-adaptive virtual topology management algorithm such as BPD could potentially enable these ad hoc players to join or drop out dynamically. Thus, it would be interesting to study how the algorithm performs in the presence of such entities.

This paper was mainly used to introduce the algorithm concepts and show it on a working example. In the future, large-scale experiments on more complex networks need to be performed. Comparison between the proposed scheme and some other node linking choices such as connecting two intermediate nodes should also be considered. Techniques for optimizing the topology based on the number of communication links and desired convergence, as well as a time-synchronized implementation of the averaging algorithm can also be investigated. 


\section*{Acknowledgment}
\footnotesize

This work is supported in part by Department of Energy's Office of Energy Efficiency and Renewable Energy (EERE) under the Solar Energy Technologies Office (SETO) Award Number DE-EE0008769.
This work was authored in part by Siemens Technology and the National Renewable Energy Laboratory, operated by Alliance for Sustainable Energy, LLC, for the U.S. Department of Energy (DOE) under Contract No. DE-AC36-08GO28308. 
Any opinions, findings, and conclusions or recommendations expressed in this material are those of the author(s) and do not necessarily reflect the views of funding agency.

\bibliographystyle{IEEEtran}
\bibliography{references}

\end{document}